\documentclass[a4paper,fleqn,usenatbib]{mnras}

\usepackage{newtxtext,newtxmath}
\usepackage[T1]{fontenc}
\usepackage{ae,aecompl}

\usepackage{graphicx}	% Including figure files
\usepackage{amsmath}	% Advanced maths commands
\usepackage{amssymb}	% Extra maths symbols

\title[Circular polarimetry of SLSNe]{Testing the magnetar scenario for superluminous supernovae with circular polarimetry}

\author[A. Cikota et al.]{Aleksandar Cikota$^{1}$\thanks{E-mail: acikota@eso.org},
Giorgos Leloudas$^{2}$,
Mattia Bulla$^{3}$,
Cosimo Inserra$^{4}$,
\newauthor Ting-Wan~Chen$^5$\thanks{Alexander von Humboldt Fellow},
Jason Spyromilio$^{1}$,
Ferdinando Patat$^{1}$,
Zach Cano$^{6}$,
\newauthor Stefan Cikota$^{7,8}$,
Michael W. Coughlin$^{9}$,
Erkki Kankare$^{10}$,
Thomas B. Lowe$^{11}$,
\newauthor Justyn R. Maund$^{12}$,
Armin Rest$^{13,14}$, 
Stephen J. Smartt$^{10}$,
Ken W. Smith$^{10}$,
\newauthor Richard J. Wainscoat$^{11}$,
David R. Young$^{10}$
\\
$^{1}$European Southern Observatory, Karl-Schwarzschild-Str. 2, 85748 Garching b. M\"{u}nchen, Germany\\
$^{2}$Dark Cosmology centre, Niels Bohr Institute, University of Copenhagen, Juliane Maries vej 30, 2100 Copenhagen, Denmark\\
$^{3}$Oskar Klein Centre, Department of Physics, Stockholm University, SE 106 91 Stockholm, Sweden\\
$^{4}$Department of Physics \& Astronomy, University of Southampton, Southampton, Hampshire, SO17 1BJ, UK\\
$^{5}$Max-Planck-Institut f\"ur Extraterrestrische Physik, Giessenbachstra\ss e 1, 85748, Garching, Germany\\
$^{6}$Instituto de Astrof\'isica de Andaluc\'ia (IAA-CSIC), Glorieta de la Astronom\'ia s/n, E-18008, Granada, Spain\\
$^{7}$University of Zagreb, Faculty of Electrical Engineering and Computing, Department of Applied Physics, Unska 3, 10000 Zagreb, Croatia\\
$^{8}$Ru{\dj}er Bo\v{s}kovi\'{c} Institute, Bijeni\v{c}ka cesta 54, 10000 Zagreb, Croatia \\
$^{9}$Division of Physics, Math, and Astronomy, California Institute of Technology, Pasadena, CA 91125, USA\\
$^{10}$Astrophysics Research Centre, School of Mathematics and Physics, Queens University Belfast, Belfast BT7 1NN, UK\\
$^{11}$Institute for Astronomy, University of Hawaii, 2680 Woodlawn Drive, Honolulu, HI 96822, USA\\
$^{12}$Department of Physics and Astronomy, University of Sheffield, Hicks Building, Hounsfield Road, Sheffield S3 7RH, U.K.\\
$^{13}$Space Telescope Science Institute, 3700 San Martin Drive, Baltimore, MD 21218, USA\\
$^{14}$Department of Physics and Astronomy, The Johns Hopkins University, 3400 North Charles Street, Baltimore, MD 21218, USA
}

\date{Accepted XXX. Received YYY; in original form ZZZ}

\pubyear{2017}

\begin{document}
\label{firstpage}
\pagerange{\pageref{firstpage}--\pageref{lastpage}}
\maketitle

% Abstract of the paper
\begin{abstract}
Superluminous supernovae (SLSNe) are at least $\sim$5 times more luminous than common supernovae (SNe). Especially hydrogen-poor SLSN-I are difficult to explain with conventional powering mechanisms. One possible scenario that might explain such luminosities is that SLSNe-I are powered by an internal engine, such as a magnetar or an accreting black hole. Strong magnetic fields or collimated jets can circularly polarize light. In this work, we measured circular polarization of two SLSNe-I with the FOcal Reducer and low dispersion Spectrograph (FORS2) mounted at the ESO's Very Large Telescope (VLT). PS17bek, a fast evolving SLSN-I, was observed around peak, while OGLE16dmu, a slowly evolving SLSN-I, was observed 100 days after maximum. Neither SLSN shows evidence of circularly polarized light, however, these non-detections do not rule out the magnetar scenario as the powering engine for SLSNe-I. We calculate the strength of the magnetic field and the expected circular polarization as a function of distance from the magnetar, which decreases very fast. Additionally, we observed no significant linear polarization for PS17bek at four epochs, suggesting that the photosphere near peak is close to spherical symmetry.

\end{abstract}

% Select between one and six entries from the list of approved keywords.
% Don't make up new ones.

%%%%%%%%%%%%%%%%%%%%%%%%%%%%%%%%%%%%%%%%%%%%%%%%%%

%%%%%%%%%%%%%%%%% BODY OF PAPER %%%%%%%%%%%%%%%%%%

\section{Introduction}

Superluminous supernovae (SLSNe) may include a few remaining examples of deaths of extremely massive stars that in the early universe may have played an important role for re-ionisation of the Universe, and are therefore an important class of objects to understand. They are extremely bright, as the name would imply, and powering such a luminous display is a challenge.
Peak luminosities of SLSNe are greater by a factor of $\sim$5 than peak luminosities of type Ia supernovae, and $\sim$10-100 times greater than broad-lined type Ic and normal stripped envelope supernovae. They are separated into two classes: the hydrogen poor SLSN-I, which have quite featureless early spectra; and hydrogen-rich SLSN-II, which are thought to occur within a thick hydrogen shell and are therefore difficult to investigate \citep{2012Sci...337..927G}.

\citet{2007Natur.450..390W} suggest that collisions between shells of matter ejected by massive stars, that undergo an interior instability arising from the production of electron-positron pairs might explain such luminous SLSNe-I \citep[see also][]{2016arXiv160808939W} or a pair-instability explosion of a very massive star \citep[with a core of $\geq50M_\odot$, e.g.][]{2009Natur.462..624G,2013MNRAS.428.3227D}. The luminosity may also be produced by interaction between the ejecta and H-poor circumstellar material \citep{2012ApJ...746..121C,2017ApJ...835...58V, 2016ApJ...829...17S}. 

Another possibility is that SLSNe-I are powered by an internal engine, such as a magnetar \citep{2010ApJ...717..245K,2010ApJ...719L.204W,2013ApJ...770..128I,2013Natur.502..346N,2015MNRAS.452.1567C} or an accreting black hole \citep{2013ApJ...772...30D}. \citet{2010ApJ...717..245K} have shown that energy deposited into an expanding supernova remnant by a highly magnetic (B $\sim$ 5 $\times$ $10^{14}$ G) fast spinning neutron star can substantially contribute to the SLSN luminosity and explain the brightest events ever seen. They calculated that magnetars with initial spin periods $<$ 30 ms can reach a peak luminosity of $10^{42}$-$10^{45}$ erg $s^{-1}$ ($M_{Bol}$=-16.3 to -23.8 mag), because of the rotational energy deposition from magnetar spin-down.

In this work, we first time undertake circular polarimetry of Superluminous Supernovae in the visible part of the spectrum. We aim to test the magnetar scenario using circular polarimetry. Our hypothesis is that if there is a strong magnetic field, we would expect to observe circularly polarized light, attributed to the monotonic gray-body magnetoemissivity which has been theoretically predicted by \citet{1970ApJ...162..169K}, and demonstrated in the laboratory.
The challenge for the magnetar observations is that the energy from the magnetar is reprocessed by the ejecta so that the bulk of the luminosity is arising from thermal processes (as is manifest in the spectra). In the thermalisation process the polarization of the original light is destroyed, however, the magnetar's magnetic field will remain. 

Circular polarization has already been observed in white dwarfs with strong magnetic fields. For instance, \citet{1970ApJ...161L..77K}, and \citet{1972ApJ...171L..11A} observed strong circular polarization, 1-3$\%$, in visible light, and 8.5-15$\%$ in the infrared \citep{1970ApJ...162L..67K} of Grw+70$^{\circ}$8247. For this white dwarf they estimate a mean projected $B$ field of 1 $\times$ $10^7$ G.

Another possible origin of circularly polarized light may be an electron pitch-angle anisotropy in a relativistic jet, for instance from an accreting black hole, as suggested by \citet{2014Natur.509..201W}. They  observed circular polarization in the afterglow of Gamma-ray burst 121024A, which are believed to be powered by a collimated relativistic jet from an accreting black hole.

In section $\S$~2 we describe the targets and observations, in $\S$~3 the methods, in $\S$~4 we show the results, which we discuss in $\S$~5, and the summary and conclusions are presented in section $\S$~6.

\section{Targets and observations}

We obtained circular polarimetry of two SLSNe-I at single epochs: OGLE16dmu at 101.3 days past peak (rest frame), and PS17bek at peak brightness. Additionally, we obtained linear polarimetry of PS17bek at four different epochs ($-4.0, +2.8, +13.4$ and $+21.0$ days relative to peak brightness in rest frame).

All observations in this study were acquired with the FOcal Reducer and low dispersion
Spectrograph \citep[FORS2,][]{1967PASP...79..136A,1998Msngr..94....1A,FORS2manual} mounted at the Cassegrain focus of the UT1 Very Large Telescope (VLT), under the ESO program ID 098.D-0532(A), using the MIT CCD chip. 
The observations were obtained in the imaging polarimetry mode (IPOL). Circular polarimetry was obtained, without any filters, with two different quarter-wave retarder plate (QWP) angles of $\theta=\pm45^\circ$, but in two different rotations of the instrument ($0^\circ$ and $90^\circ$) in order to remove possible cross-talks between linear and circular polarization \citep{2009PASP..121..993B}.

Linear polarimetry of PS17bek was obtained through the V$\_$HIGH FORS2 standard filter ($\lambda_0$ = 555 nm, FWHM = 123.2 nm), at four half-wave retarder plate (HWP) angles ($0, 22.5, 45, 67.5^\circ$). 

A observation log is given in Table~\ref{tb_obs}.

\subsection{OGLE16dmu}

\begin{figure}
\includegraphics[trim=0mm 0mm 0mm 0mm, width=8.5cm, clip=true]{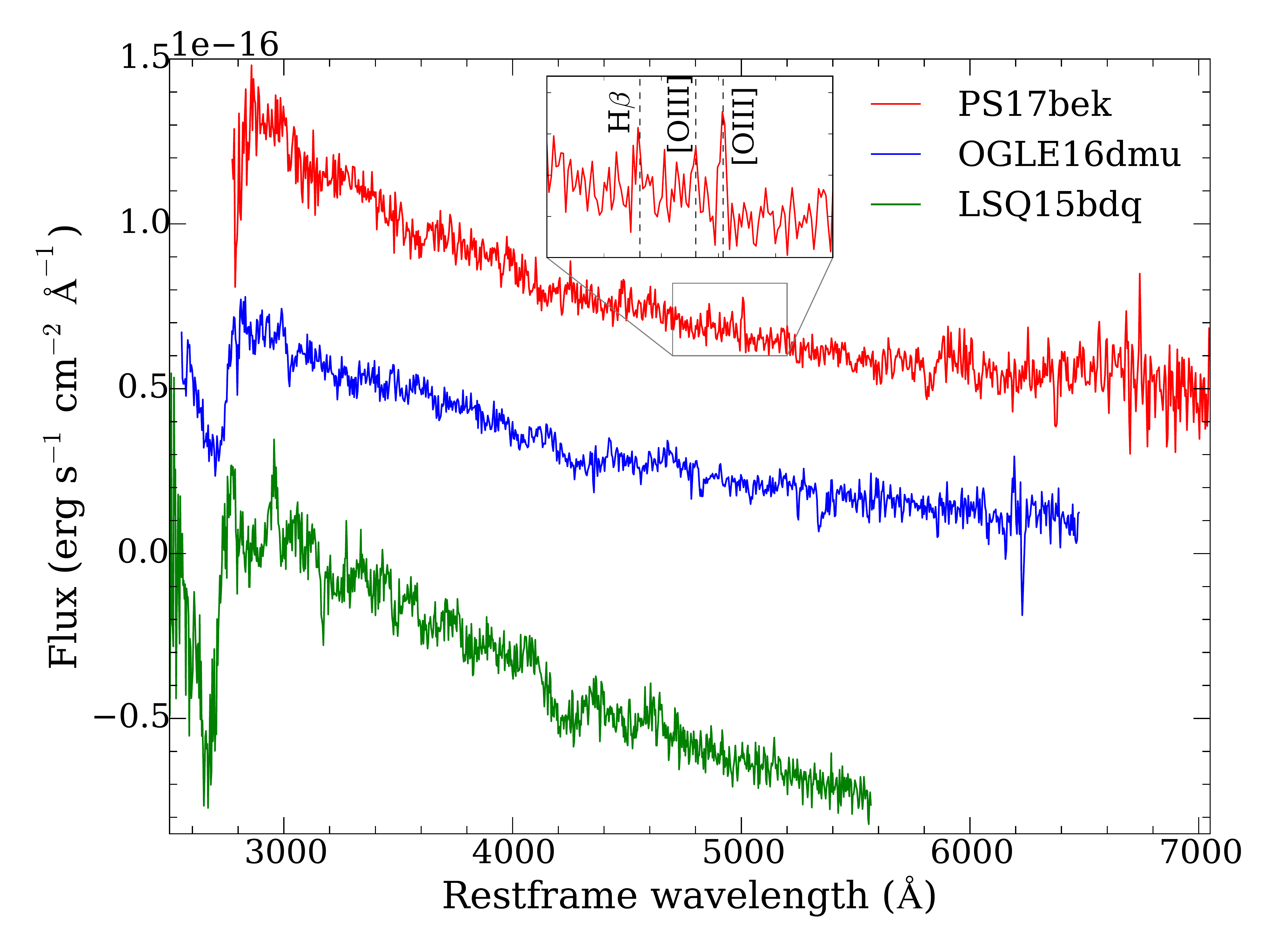}
\vspace{-6mm}
\caption{PESSTO classification spectra of OGLE16dmu (middle blue spectrum) and PS17bek (top red spectrum), compared to LSQ14bdq  (bottom green spectrum, \citealt{2015ApJ...807L..18N}). The inset shows the [OIII] and H$\beta$ emission lines in the spectrum of PS17bek, used for the redshift determination. PS17bek and LSQ14bdq have been plotted with an constant offset of +4$\times$10$^{-17}$, and -1$\times$10$^{-16}$ respectively.}
\label{fig:classificationspectra}
\end{figure}

OGLE16dmu was discovered on September 23, 2016 (MJD 57654.84) \citep{2016ATel.9543....1W}, and classified as a SLSN-I. The classification spectrum is shown in Fig.~\ref{fig:classificationspectra}. It is apparently hostless at a redshift z$\sim$0.426 \citep{2016ATel.9542....1P}. 
% 2410.2 Mpc
From GROND observations (Chen et al., in preparation), we determined an apparent magnitude at peak of m$_{\rm r}$ = 19.41  mag in November 11, 2016 (MJD 57698.41). The total Galactic reddening in the direction of OGLE16dmu is $E(B-V)$ = 0.03 mag \citep{2011ApJ...737..103S}, which corresponds to A$_r$ $\sim$ 0.07 mag assuming a \citet{Fitzpatrick1999PASP..111...63F} extinction law and $R_V$=3.1. The Galactic reddening-corrected absolute brightness is M$_{\rm r}$ = -22.2  mag.\footnote{we assume a flat universe with $H_0=67.8$ km s$^{-1}$ Mpc$^{-1}$ and $\Omega_M=0.308$ \citep{2016A&A...594A..13P}.}

From the rest-frame light curve, we estimate the rate of decline at 30 days past maximum \citep{2014ApJ...796...87I} to be DM$_{30} \sim$ 0.22 mag. Alternatively, using the metric described in \citet{2015MNRAS.452.3869N} (the time to reach from maximum light, f$_{max}$, to f$_{max}$/e), we estimate $\tau_{dec} \sim$ 70.6 days. Thus, this is a bright and slowly-evolving SLSN-I, similar to PTF12dam or SN~2015bn.

\subsection{PS17bek}

PS17bek is a SLSN-I at z=0.30992 $\pm$ 0.0003 (see Fig.~\ref{fig:classificationspectra}, PESSTO classification). 

It was discovered at $\alpha$=10$^{\rm h}$47$^{\rm m}$41.90$^{\rm s}$ and $\delta$= +26$^\circ$50'06.0'' on MJD=57802.4 (2017  Feb 18.4) and it is possibly associated to the galaxy GALEXMSC J104742.19+265006.8. The object was discovered when this region of sky was observed by Pan-STARRS \citep{2016arXiv161205560C,2016MNRAS.462.4094S} in response to a possible low significance gravitational wave signal provided by LIGO-Virgo \citep{2016ApJ...826L..13A}, but this transient was not considered related to that event.
As part of the Public ESO Spectroscopic Survey for Transient Objects (PESSTO), we took a a classification 
spectrum (see \citealt{2015A&A...579A..40S} for details of the instrumentation, calibration an data access). 

%It was discovered at $\alpha$=10:47:41.90 and $\delta$= +26:50:06.0 and it is possibly associated to the galaxy GALEXMSC\, J104742.19+265006.8. 
%dl=1661.9 Mpc

%This region of sky was observed by Pan-STARRS in response to a possible low significance gravitational wave signal provided to us by LIGO-Virgo \citep{2016ApJ...826L..13A}, but this transient is not considered related to that event. 

We determined an apparent magnitude at peak of m$_{\rm r}$ = 19.8 mag (Cano et al., in preparation) at MJD = 57814.58 days. The Galactic reddening in the direction of PS17bek is $E(B-V)$ = 0.03 mag \citep{2011ApJ...737..103S}, which corresponds to A$_r$ $\sim$ 0.07 mag. Thus, the Galactic reddening-corrected absolute magnitude of PS17bek is M$_r$ $\sim$ -20.7 mag.

For PS17bek we estimate a decline rate of  DM$_{30} \sim$ 1.62 mag or t$_{dec} \sim$ 23 days. Thus, this is a fast-declining SLSN-I, similar to SN~2010gx or SN~2011ke. In fact, the measured decline rate implies that PS17bek is one of the fastest evolving SLSNe-I \citep[see][]{2018ApJ...854..175I}. 
Starting from \citet{2012Sci...337..927G}, it remains an unresolved issue if H-poor SLSNe can be divided into more subclasses (e.g. Type I/Type R or fast/slow) and whether this division has physical implications \citep{2018ApJ...854..175I, 2017arXiv170801623D, 2018ApJ...855....2Q}. Irrespective, it remains an advantage that our experiment probes representative SLSNe from both sub-classes.

\begin{table*}
\caption{Observations log}
\label{tb_obs}
%\vspace{-0.3cm}
\centering
\begin{tabular}{lllllccc}
\hline\hline
Name   & UT Date and Time & Filter & $\lambda/2$-plate   & $\lambda/4$-plate    & Wollaston   &  Exposure  & Seeing \\
        &         &        &  angle [$^{\circ}$] &  angle [$^{\circ}$] & angle [$^{\circ}$]  & [s]  &     ['']   \\
\hline  
PS17bek   &  2017-02-25 05:48:09	 &  None	  & \dots & 315 & 0	  	 & 200	 & 0.61 \\
PS17bek   &  2017-02-25 05:53:59	 &  None	  & 	\dots& 45  & 0		 & 200	 & 0.62 \\
PS17bek   &  2017-02-25 06:23:44	 &  None	  & \dots & 405 & 90		 & 200	 & 0.67 \\
PS17bek   &  2017-02-25 06:28:02	 &  None	  & 	\dots & 135 & 90	     & 200	 & 0.61 \\
&&&&&&&\\
PS17bek   &  2017-02-25 06:42:04	 & v\_HIGH & 0	 & 	\dots	 & 0	 & 650	 & 0.67 \\
PS17bek   &  2017-02-25 06:53:37	 & v\_HIGH & 45	 & \dots		 & 0	 & 650	 & 0.63 \\
PS17bek   &  2017-02-25 07:05:03	 & v\_HIGH & 22.5 & \dots		 & 0	 & 650	 & 0.66 \\
PS17bek   &  2017-02-25 07:16:35	 & v\_HIGH & 67.5 & \dots		 & 0	 & 650	 & 0.55 \\
&&&&&&&\\ 
PS17bek   &  2017-03-06 05:07:26	 & v\_HIGH & 0	 & \dots		 & 0	 & 520	 & 0.71  \\
PS17bek   &  2017-03-06 05:16:48	 & v\_HIGH & 45   & \dots		 & 0	 & 700	 & 0.60  \\
PS17bek   &  2017-03-06 05:29:04	 & v\_HIGH & 22.5 & \dots		 & 0	 & 700	 & 0.59  \\
PS17bek   &  2017-03-06 05:41:26	 & v\_HIGH & 67.5 & \dots		 & 0	 & 700	 & 0.50  \\
&&&&&&&\\ 
PS17bek   &  2017-03-20 01:45:38	 & v\_HIGH & 0	 & \dots		 & 0	 & 700	 & 0.64 \\
PS17bek   &  2017-03-20 01:58:02	 & v\_HIGH & 45   & \dots		 & 0	 & 700	 & 0.69 \\
PS17bek   &  2017-03-20 02:10:17	 & v\_HIGH & 22.5 & \dots		 & 0	 & 700	 & 0.70 \\
PS17bek   &  2017-03-20 02:22:40	 & v\_HIGH & 67.5 & \dots		 & 0	 & 700	 & 0.86 \\
PS17bek   &  2017-03-20 02:36:04	 & v\_HIGH & 0	 & 	\dots	 	 & 0	 & 700	 & 0.68 \\
PS17bek   &  2017-03-20 02:48:27	 & v\_HIGH & 45   & \dots		 & 0	 & 700	 & 0.67 \\
PS17bek   &  2017-03-20 03:00:42	 & v\_HIGH & 22.5 & \dots		 & 0	 & 700	 & 0.77 \\
PS17bek   &  2017-03-20 03:13:05	 & v\_HIGH & 67.5 & \dots		 & 0	 & 700	 & 0.81 \\
&&&&&&&\\ 
PS17bek   &  2017-03-30 02:10:19	 & v\_HIGH & 0	 & 	\dots	 	 & 0	 & 500	 & 0.69 \\
PS17bek   &  2017-03-30 02:19:23	 & v\_HIGH & 45   & \dots		 & 0	 & 500	 & 0.72 \\
PS17bek   &  2017-03-30 02:28:18	 & v\_HIGH & 22.5 & \dots		 & 0	 & 500	 & 0.68 \\
PS17bek   &  2017-03-30 02:37:21	 & v\_HIGH & 67.5 & \dots		 & 0	 & 500	 & 0.56 \\
PS17bek   &  2017-03-30 02:46:59	 & v\_HIGH & 0    & \dots		 & 0	 & 500	 & 0.47 \\
PS17bek   &  2017-03-30 02:56:02	 & v\_HIGH & 45   & \dots		 & 0	 & 500	 & 0.54 \\
PS17bek   &  2017-03-30 03:04:57	 & v\_HIGH & 22.5 & \dots		 & 0	 & 500	 & 0.58 \\
PS17bek   &  2017-03-30 03:13:60	 & v\_HIGH & 67.5 & \dots		 & 0	 & 500	 & 0.81 \\
&&&&&&&\\ 
OGLE16dmu &  2017-03-30 23:59:36	 &  None	  & 	\dots & 315   & 0	 & 220	 & 0.74 \\
OGLE16dmu &  2017-03-31 00:04:14	 &  None  & 	\dots & 45	 & 0	 & 220	 & 0.85 \\
OGLE16dmu &  2017-03-31 00:18:00	 &  None	  & \dots	 & 405 & 90	     & 220	 & 0.84 \\
OGLE16dmu &  2017-03-31 00:22:38	 &  None	  & \dots	 & 135 & 90	     & 220	 & 0.75 \\
\hline
\end{tabular}
\end{table*}
%% DIMM  seeing
%% 098.D-0532(A)

\section{Data processing and methods}

The data consist of two science frames per exposure: the upper CHIP1 and lower CHIP2, which correspond to two mosaic-parts of the two CCD detectors. In IPOL mode, the image is split by the Wollaston prism into an ordinary (o) beam and an extra-ordinary (e) beam, and the multi-object spectroscopy (MOS) slitlets strip mask is inserted to avoid the beams overlapping. The targets were observed at the bottom of CHIP1 (upper frame), centered in the optical axis of the telescope. The bottom strip in the upper frame is the extra-ordinary beam.
The Wollaston prism is usually aligned with the north celestial meridian except when the instrument is rotated by 90$^{\circ}$ during the second sequence of circular polarimetry, when it was aligned towards East.

All frames were bias subtracted using the corresponding calibration bias frames. A flat-field correction was not performed because the flat-field effect gets canceled out,  because of the redundancy introduced by multiple HWP and QWP angles, for linear and circular polarimetry respectively \citep{FORS2manual,2006PASP..118..146P}.

To determine the polarization of our targets, we conducted aperture photometry of sources in the ordinary and extra-ordinary beams using the IRAF's DAOPHOT.PHOT package. An optimal aperture radius of $\sim$2 FWHM was used.

\subsection{Circular polarimetry}

Following the FORS2 user manual \citep{FORS2manual}, the amount of circular polarization is given as:
\begin{equation}
V =\frac{1}{2}  \left[ \left(\frac{f^o - f^e}{f^o + f^e}\right)_{\theta=45^{\circ}}    -  \left(\frac{f^o - f^e}{f^o + f^e}\right)_{\theta=-45^{\circ}}  \right]
\end{equation}
where $f^o$ and $f^e$ are the measured flux in the ordinary and extra-ordinary beam respectively, for both quarter-wave retarder plate angles of $\theta= \pm 45^{\circ}$.
The circular polarization error was calculated by error propagation of the flux errors.

To minimize a possible linear-to-circular polarization cross talk \citep{2009PASP..121..993B}, we calculate the average of the Stokes $V$ measured at two instrument position angles, $\phi$, and $\phi$+90$^{\circ}$:
\begin{equation}
P_{V} = \frac{V_{\phi}+V_{\phi+90^{\circ}}}{2} ,
\end{equation}
which leads to cancellation of the spurious signal \citep{2009PASP..121..993B}.

% Bagnulo: \citep{2009PASP..121..993B}

%\begin{equation}
%err1 =(f45o/epadu + area_45o*(stdev_45o^2) + (area_45o^2)*(stdev_45o^2)/nsky_45o[0])^0.5
%\end{equation}

%\begin{equation}
%\begin{split}
%\sigma_V = \left[ \left(\frac{f^e_{45^{\circ}}}{(f^o_{45^{\circ}}+f^e_{45^{\circ}})^2} \times err1\right)^2   +   \left(\frac{f^o_{45^{\circ}}}{(f^o_{45^{\circ}}+f^e_{45^{\circ}})^2} \times err2\right)^2   +  \right. \\
%		\left. \left(\frac{-f^e_{315^{\circ}}}{(f^o_{315^{\circ}}+f^e_{315^{\circ}})^2} \times err3\right)^2  +  \left(\frac{f^o_{315^{\circ}}}{(f^o_{315^{\circ}}+f^e_{3315^{\circ}})^2} \times err4\right)^2 \right]^{\frac{1}{2}}
%\end{split}
%\end{equation}

\subsection{Linear polarimetry}
 
The Stokes $Q$ and $U$ parameters for PS17bek and a number of field stars were derived using the standard approach, as described in \citet{2015ApJ...815L..10L}; that is via the Fourier transformation of normalized flux differences measured at four half-wave retarder plate angles of 0, 22.5, 45 and 67.5$^\circ$ \citep[see also the FORS2 manual, ][]{FORS2manual}.

\begin{figure}
\includegraphics[trim=15mm 5mm 10mm 25mm, width=9cm, clip=true]{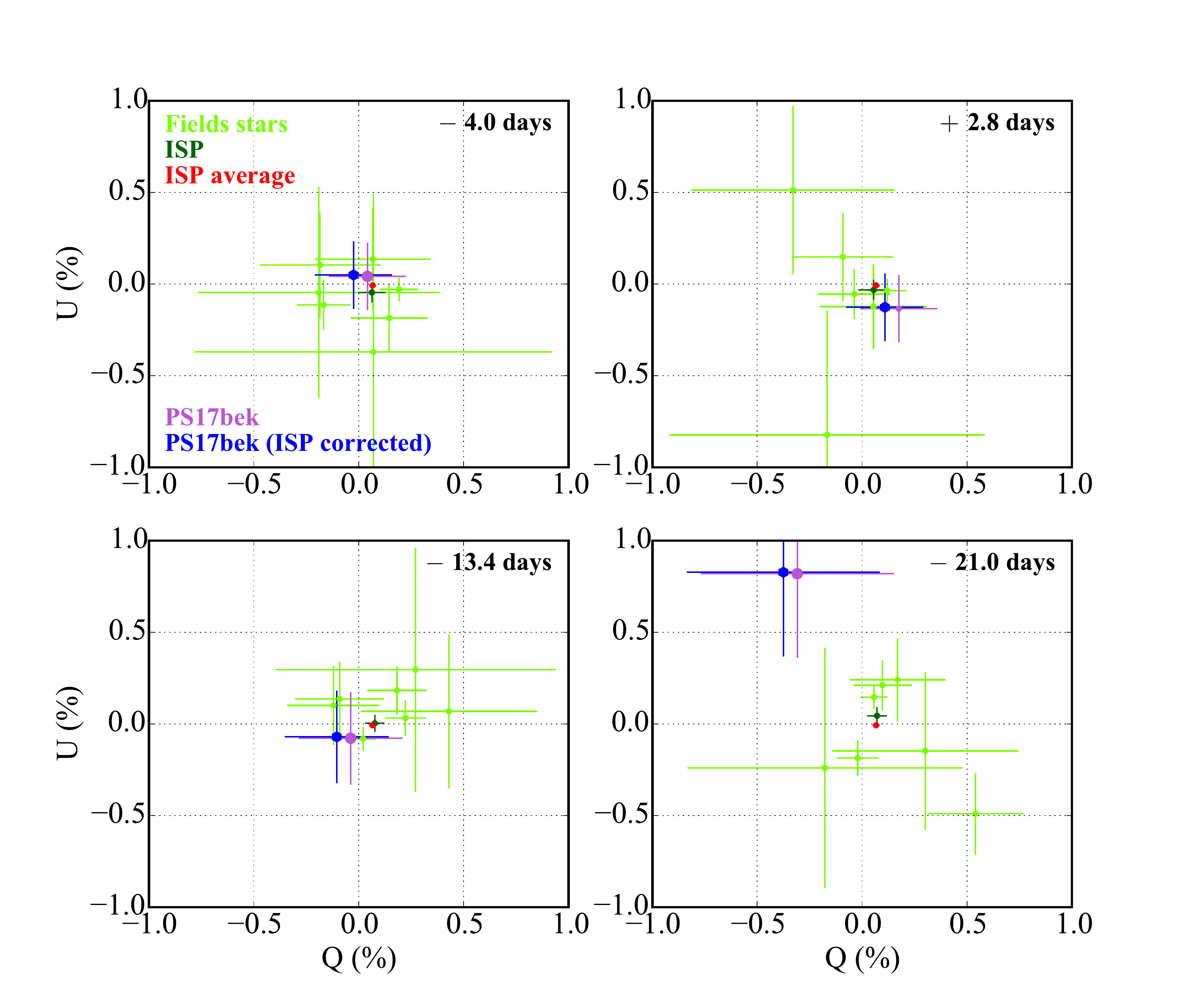}
\vspace{-5mm}
\caption{Q--U plane for all 4 epochs of PS17bek. Comparison stars are colored light green. In each panel, a dark green cross indicates the position of the ISP, calculated as the barycenter of the stars at each epoch. The red cross is the ISP averaged over all epochs, which coincides with the dark green cross in the individual epochs. The original measurement of the SN is shown in magenta and the ISP corrected value in blue.}
\label{fig:ISM_lin_PS17bek}
\end{figure}

We correct the polarization position angles of the raw measurements for the half-wave plate zero angle chromatic dependence \citep[Table 4.7 in][]{FORS2manual}, and for the instrumental polarization, which increases with distance from the optical axis \citep[Fig.~5 in][]{2006PASP..118..146P}. In addition, we used 7 field stars to determine the interstellar polarization (ISP) by calculating their barycenter in the $Q$--$U$ plane for each epoch (Fig.~\ref{fig:ISM_lin_PS17bek}). The stars give a stable and self-consistent result with time:\\
$Q_{ISP}$ = 0.066 $\pm$ 0.004 $\%$ \\
$U_{ISP}$ = -0.007 $\pm$ 0.018 $\%$.\\
Thus, P$_{ISP}$ = 0.066 $\pm$ 0.004 $\%$. This value is lower than the expected maximum interstellar polarization, p$_{max}$ = 9.0 $\times$ $E(B-V)$, determined by \citet{1975ApJ...196..261S}, using the Galactic reddening in the direction of PS17bek, $E(B-V)$ = 0.027 $\pm$ 0.004 mag \citep{2011ApJ...737..103S}.

Additionally, we do a polarization bias correction, following \citet{2006PASP..118..146P}.

%using the following equation, given in %\citet{espinosa1997instrumentation}:
%\begin{equation}
%P \sim  P_{obs} \sqrt{1- \left( \frac{\sigma_P}{P_{obs}} \right)^2}
%\end{equation}

\section{Results}

We undertook circular polarimetry for two SLSNe-I: OGLE16dmu 101.3 days after peak brightness (in rest frame), and PS17bek 4.0 days before peak brightness. 

The circular polarization of both SLSNe is consistent with zero. We measured a circular polarization of $P_V$=-0.55 $\pm$ 1.31 $\%$ for OGLE16dmu, and $P_V$=-0.21 $\pm$ 0.18 $\%$ for PS17bek. The results are summarized in Table~\ref{tab1}. The signal-to-noise ratio of PS17bek observed at different instrument rotation angles $\phi$ of 0 and 90 degrees is S/N$\sim$272 and $\sim$172 respectively, while for OGLE16dmu S/N$\sim$62 (at $\phi$=0 deg) and $\sim$59 (at $\phi$=90 deg), which explains the large uncertainties of the calculated polarization\footnote{The absolute error of P is related to the signal-to-noise ratio as $\sigma_P$ = $\frac{1}{\sqrt{N/2} \, SNR}$, where N is the number of waveplate angles used \citep{2006PASP..118..146P}.}.

\begin{figure*}
\centering
\includegraphics[trim=0mm 0mm 0mm 0mm, width=7.73 cm, clip=true]{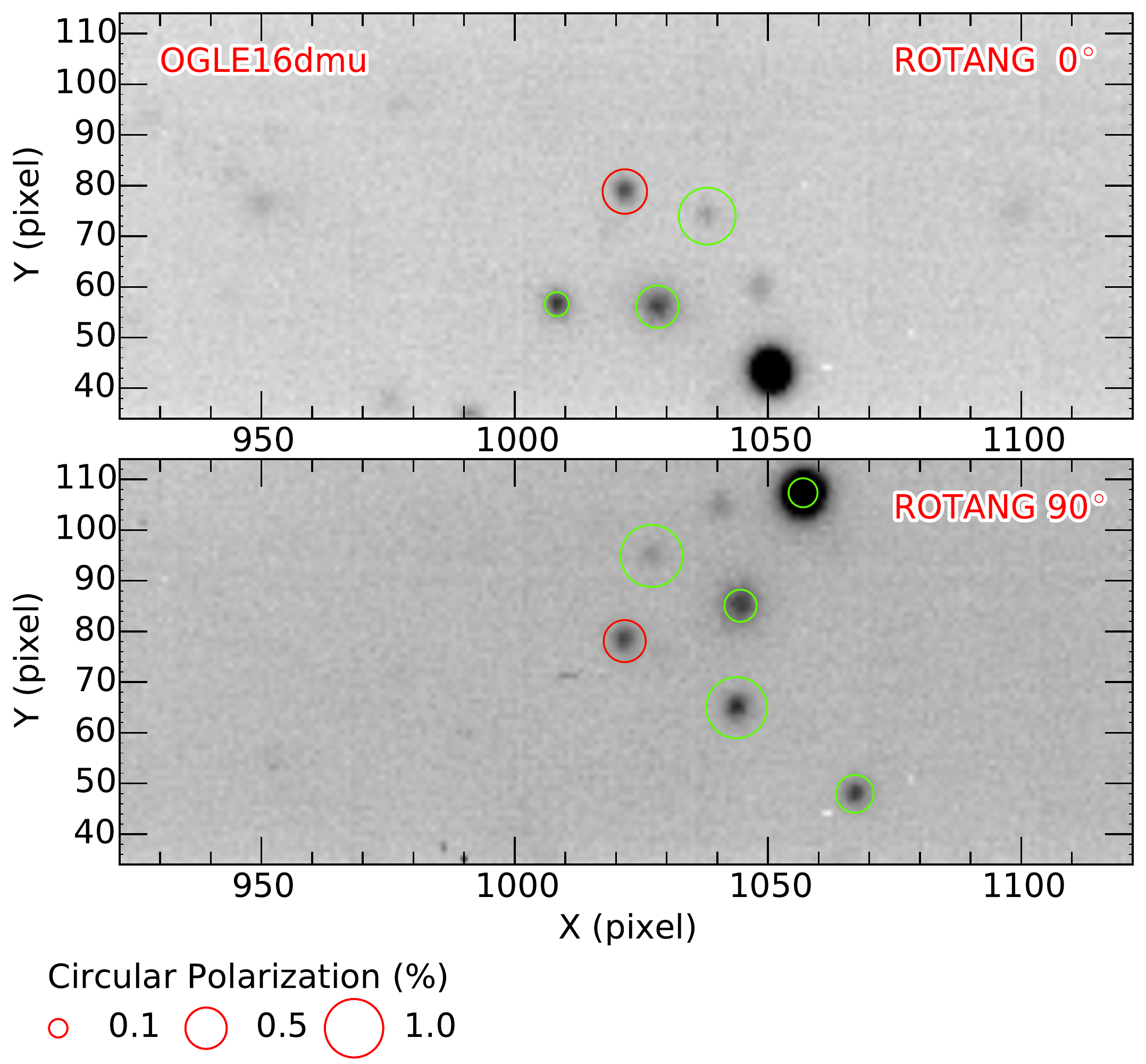}
\hspace{6mm}
\includegraphics[trim=0mm 0mm 0mm 0mm, width=9.1 cm, clip=true]{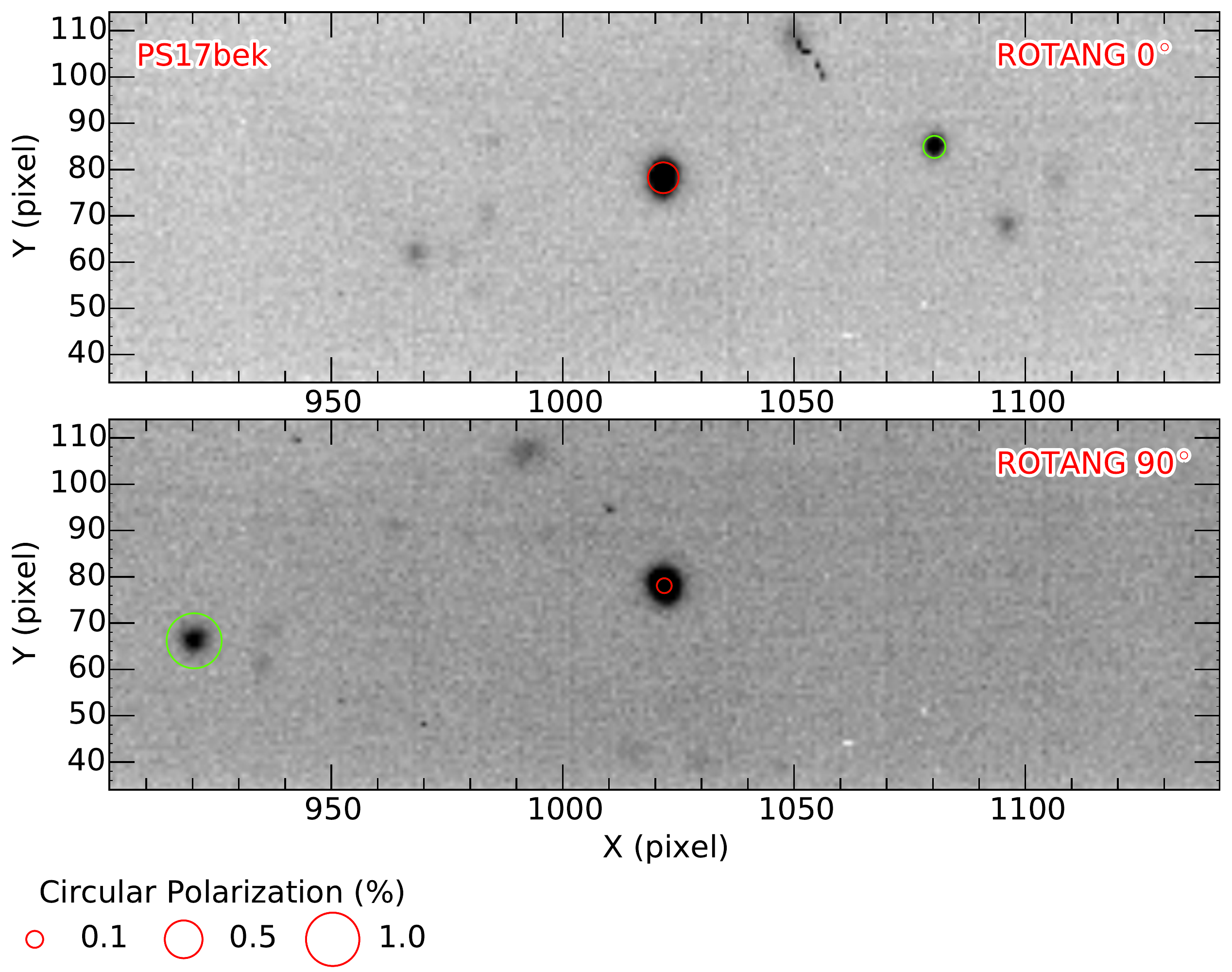}
\vspace{-1mm}
\caption{Sections of the ordinary beams for single imaging polarimetry exposures for OGLE16dmu (left) and PS17bek (right). The top and bottom panels are exposures taken with the instrument rotated by 0$^{\circ}$ and 90$^{\circ}$ respectively. The red circles mark the targets, while green circles mark comparison stars in the field. The radii of the circles correspond to the absolute circular polarization, as indicated in the legend.}%, as indicated in the legend.}
\label{fig:OGLE16dmu_and_PS17bek}
\end{figure*}

Figure~\ref{fig:OGLE16dmu_and_PS17bek} shows a section of the FORS2 imaging polarimetry field for OGLE16dmu and PS17bek taken with different instrument position angles. It is shown that the measured polarization of our targets is consistent with the polarization of field stars which are expected to be unpolarized. Furthermore, the fainter sources with lower S/N have larger polarization values, but also higher uncertainties.

\begin{figure}
\includegraphics[trim=0mm 0mm 0mm 19mm, width=9cm, clip=true]{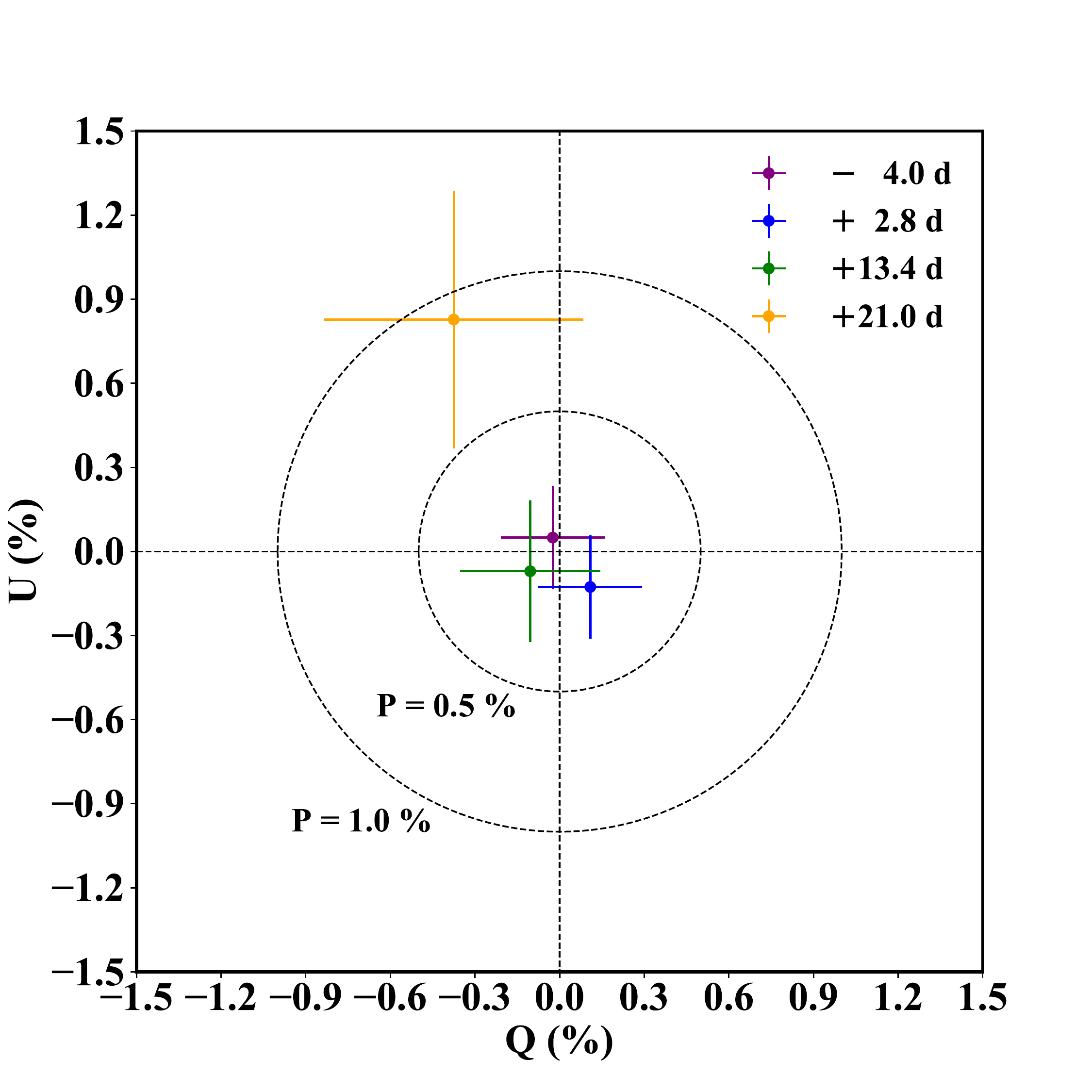}
\vspace{-5mm}
\caption{Stokes $Q$--$U$ plane for PS17bek observed at four epochs. The different colors indicate different epochs: -4.0 (purple), +2.8 (blue), +13.4 (green) and +21.0 (yellow) days relative to peak brightness. The dashed concentric circles of equal polarization have a radius of 0.5 $\%$ and 1.0 $\%$ respectively.}
\label{fig:lin_PS17bek}
\end{figure}

The ISP-corrected linear polarization measurements of PS17bek are given in Table~\ref{tab2}, and shown in Figure~\ref{fig:lin_PS17bek}. At least for the first 3 epochs (-4, +2.8, +13.4 relative to peak brightness) the linear polarization of the SLSN is very similar to one of the field stars and consistent with zero in $Q$ and $U$. Thus, there is no significant linear polarization at these phases. The fourth epoch (21.0 days past maximum brightness) might indicate a larger polarization ($\sim$0.8 $\%$) but the result is not highly significant. The SNR at the last phase is 154 (since the SN has faded), which is significantly lower than at -4 days (SNR$\sim$384), +2.8 days (SNR$\sim$384) and +13.4 days (SNR$\sim$282) relative to peak brightness.
Considering that the uncertainty of the last phase is $\sim$0.5$\%$, this is a 2$\sigma$ result.

\begin{table*}
%\centering
\caption{Circular polarimetry results}
\label{tab1}
%\vspace{-0.3cm}
%\centering
\begin{tabular}{lcccc}
\hline\hline
SLSN  & Phase & V$_{0^{\circ}}$   ($\%$) &  V$_{90^{\circ}}$ ($\%$)   &  $P_V$  ($\%$)  \\
\hline
PS17bek   & -4.0 d  &-0.33 $\pm$ 0.25  & -0.08 $\pm$ 0.27  & -0.21 $\pm$ 0.18 \\
OGLE16dmu & +101.3 d & -0.58  $\pm$   1.30  & -0.52  $\pm$  2.28 & -0.55 $\pm$ 1.31 \\
\hline
\end{tabular}
\end{table*}

\begin{table}
%\center
\caption{ISP corrected linear polarimetry results for PS17bek}
\label{tab2}
%\vspace{-0.3cm}
\centering
\begin{tabular}{lcccc}
\hline\hline
Phase     &  Q ($\%$)  & U ($\%$) &   $P^*$ ($\%$) &   $\phi$ ($^{\circ}$) \\
\hline
-4.0 & -0.02 $\pm$ 0.18 & 0.05 $\pm$ 0.18 & 0.0 $\pm$ 0.18 & 56.3 $\pm$ 97.1\\
+2.8 & 0.1 $\pm$ 0.18 & -0.13 $\pm$ 0.18 & 0.0 $\pm$ 0.18 & -26.5 $\pm$ 31.6\\
+13.4& -0.11 $\pm$ 0.25 & -0.06 $\pm$ 0.25 & 0.0 $\pm$ 0.25 & -74.6 $\pm$ 56.8\\
+21.0& -0.32 $\pm$ 0.46 & 0.85 $\pm$ 0.46 & 0.19 $\pm$ 0.46 & 55.4 $\pm$ 14.5\\
%
%
% chromatism uncorrected
% -4.0  &  -0.02 $\pm$ 0.18 &  0.05 $\pm$  0.19  & 0.06  $\pm$  0.19 &   58.1 $\pm$  95.8\\
%  2.8  &   0.11 $\pm$ 0.18 & -0.13 $\pm$  0.18   & 0.17  $\pm$  0.18 &  -24.7 $\pm$  31.7\\
% 13.4  &  -0.1 $\pm$ 0.25 & -0.07 $\pm$  0.25  & 0.13  $\pm$  0.25 &  -73.0 $\pm$  57.3\\
% 21.0  &  -0.38 $\pm$ 0.46 &  0.83 $\pm$  0.46  & 0.91  $\pm$  0.46 &   57.2 $\pm$  14.5\\
\hline
\multicolumn{5}{l}{$^*$ polarization-bias corrected}\\
\end{tabular}
\end{table}

\section{Discussion}

\subsection{Circular polarimetry of OGLE16dmu and PS17bek}

In the magnetar scenario, a rapidly rotating magnetar is born during a core-collapse SN explosion. The explosion ejects many solar masses of material, which expands while the magnetar spins down. The spin-down injects $\sim$10$^{51}$ ergs into the ejected material, that has since expanded to a distance of $\sim$100 AU, and heats it up, which then radiates the energy away \citep{2010ApJ...719L.204W, 2010ApJ...717..245K,2013ApJ...770..128I,Smith2015}.

The idea behind observing a target at early phases was to possibly detect an imprint of the strong magnetic field in the ejected material, while the aim of observing a target at late phases was to observe emitted light originating from the photosphere which moves inwards with time, closer to the magnetar, as the ejecta expands and becomes transparent.

\citet{1970ApJ...162..169K} predicted that a "gray-body" model in a magnetic field will emit a fraction of circularly polarized light. The degree of polarization, q, is proportional to the emitting wavelength, $\lambda$, and the strength of the magnetic field, $B$ \citep[see Eq. 7 and 16 in][]{1970ApJ...162..169K}, and is given by:
\begin{equation}
\label{eq_qpol}
q(\lambda) \simeq - \frac{\lambda eB}{4 \pi m c}, 
\end{equation}
where $e$ and $m$ are the electron's charge and mass, respectively, and $c$ is speed of light.

However, since the magnetic field is decreasing with distance, proportional to 1/distance$^3$, the polarization will drop very quickly. 
Assuming a magnetic field $B_0$ at the surface of a magnetar with radius $R_0$, the maximum magnetic field decreases as a function of distance, r, as following:
\begin{equation}
B(r) = B_0 \left(  \frac{R_0}{r} \right ) ^3 .
\end{equation}

\begin{figure}
\includegraphics[trim=0mm 0mm 0mm 0mm, width=9cm, clip=true]{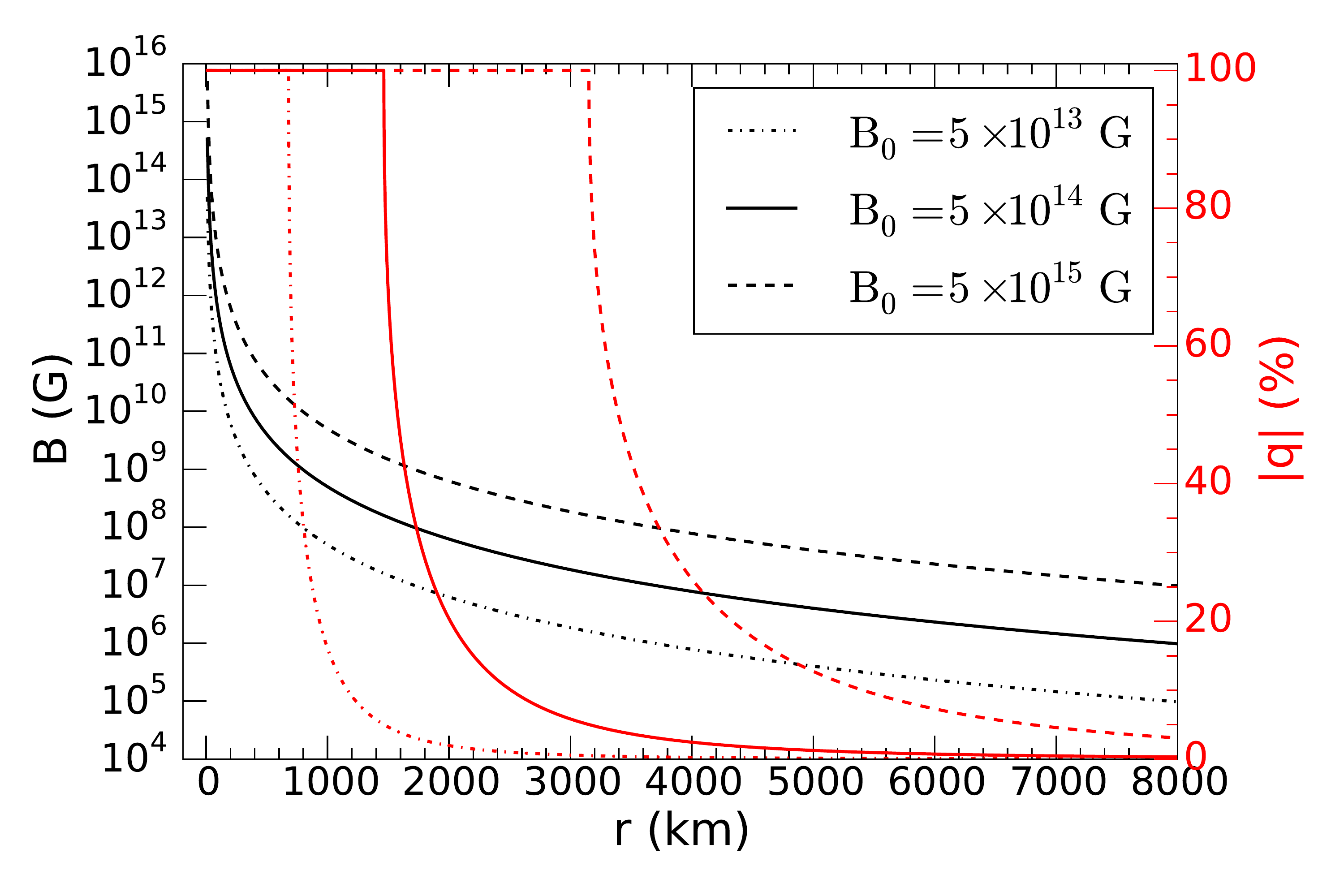}
\vspace{-7mm}
\caption{Maximum magnetic field strength (black lines) and absolute circular polarization, q (red lines), in the optical ($\lambda$=0.67 $\mu$m) as a function of distance, r, for three different initial surface magnetic field strengths, $B_0$, at $R_0$=10 km.}
\label{fig:B-q}
\end{figure}

Figure~\ref{fig:B-q} shows the magnetic field, B, and the circular polarization attributed to gray-body magnetoemissivity, q, as a function of distance, calculated in the optical ($\lambda$ = 0.67 $\mu m$), for three different surface magnetic strengths, $B_0$, for a magnetar of radius $R_0$ = 10 km.

For example, assuming a surface magnetic field strength of $B_0$=5 $\times$ 10$^{15}$ G, the magnetic field strength drops to 4$\times$10$^4$ G at a distance of only 5 $\times$10$^4$ km. The degree of polarization produced by gray-body magnetoemissivity at that distance is q$\sim$ 0.01 $\%$, which is beyond our detection capabilities. 

Furthermore, our observations were taken without any filter in order to achieve a high SNR in a reasonable time, while the absolute degree of circular polarization produced by gray-body magnetoemissivity increases with wavelength (see Eq.~\ref{eq_qpol}). Therefore, it is generally recommended to observe circular polarization at infrared wavelengths.

Despite a non-detection of circular polarization in SLSN-I, the magnetar scenario cannot be excluded as the internal engine of SLSNe, because in order to observe circularly polarized light attributed to gray-body magnetoemissivity, it is necessary that the light is emitted within strong magnetic fields, close to the magnetar, which is not the case in the magnetar scenario as described e.g. by \citet{2010ApJ...717..245K}, in contrast to the observed circular polarization in white dwarfs \citep[e.g.][]{1971ApJ...164L..17K,1973ApJ...180L.123R}, where the observed light is emitted from the white dwarf's surface.

Another possibility for the lack of observed circular polarization is that OGLE16dmu and PS17bek are not driven by an internal engine at all. For instance, other possible scenarios that could explain such a high luminosity is a pair-instability supernova \citep[PISN, e.g.][]{2007Natur.450..390W,2009Natur.462..624G,2013MNRAS.428.3227D,2016arXiv160808939W}, or a normal SN explosion interacting with circumstellar shells \citep[e.g.][]{2012ApJ...746..121C,2017ApJ...835...58V, 2016ApJ...829...17S}. In case of a PISN, which requires high amounts of $^{56}$Ni to explain the luminosity, the light curves are expected to evolve slowly, which likely rules out this scenario for PS17bek that has one of the fastest evolving light curves (Chen et al., in preparation). However, it is beyond the scope of this short paper to analyze the light curves for those SLSNe.

\subsection{Linear polarimetry of PS17bek}

Intrinsic linear polarization of SNe is a measure of the supernova's photosphere departure from spherical symmetry projected on the sky. If the projection of the photosphere is not symmetric, more photons will be scattered by electrons along the photosphere's major axis than along the minor axis, which will produce net-polarization in the continuum \citep[see e.g.][]{1991A&A...246..481H,2003ApJ...593..788K,2015MNRAS.450..967B}.

Because SLSNe are faint, and thus it is hard to undertake polarimetry which requires high SNR, only a few SLSNe have been studied using polarimetry \citep{2015ApJ...815L..10L,2016ApJ...831...79I, 2018MNRAS.475.1046I,2017ApJ...837L..14L,2018ApJ...853...57B}. 

LSQ14mo, also a fast-declining SLSN-I (as PS17bek), did not show evidence for significant polarization or polarization evolution  from $-7$ and up to $+19$ days with respect to maximum \citep{2015ApJ...815L..10L}.
In the contrary, the slowly-evolving SN~2015bn did show an increase in polarization with time, that was attributed to the photosphere receding to inner layers of the explosion that are more asymmetric. 
\citet{2016ApJ...831...79I} obtained the first spectropolarimetric observations of a SLSN-I, at $-24$ and $+28$ days, further showing that the geometry was consistent with an axisymmetric configuration (that could be consistent with a magnetar scenario). The polarization increase was confirmed by 
\citet{2017ApJ...837L..14L} who obtained multi-epoch imaging polarimetry between $-20$ and $+46$ days, showing that the increase was coincident with changes in the optical spectrum.

The result obtained for PS17bek is fairly consistent with the picture obtained from previous events.
Similarly to the other SLSNe-I, observed around peak, no significant polarization is detected. Our last observation (at $+21$ days) could be consistent with an increase in polarization but the significance of this result is below 2$\sigma$. Either fast-evolving SLSNe (PS17bek and LSQ14mo) follow a different geometrical evolution than slowly-evolving SLSNe, or simply the available data, due to a combination of low SNR and lack of data at late phases, are not able to significantly detect an increase in polarization.

\section{Summary and conclusions}

In this work, we investigated circular polarization of two hydrogen poor superluminous supernovae for the first time, using FORS2 at the VLT. Our main results can be summarized as follows:

\begin{enumerate}
\item OGLE16dmu is a slowly evolving hydrogen poor SLSN. We undertook circular imaging polarimetry at +101.3 days past peak (in rest frame r band) and found no evidence of circular polarization.
\item PS17bek is a fast evolving SLSN-I. We undertook circular polarimetry at -4.0 days relative to the peak brightness (in rest frame r band), and found no evidence of circular polarization. 
\item Additionally, PS17bek was observed in linear polarimetry mode at four phases (-4.0, +2.8, +13.4 and +21.0 days), and shows no significant linear polarization.
\item We cannot exclude the magnetar scenario because of a non-detection of circular polarization, which, due to the rapid decrease in the strength of the magnetic with distance, would only be detectable at small radii close to the surface of the magnetar.
\item We note that future attempts to measure the strength of magnetic fields using circular polarimetry should be made in the infrared, where the expected degree of circular polarization produced by gray-body magnetoemissivity is higher. 
\item It is not likely that we will observe circular polarization produced by gray-body magnetoemissivity, because (assuming the magnetar scenario) the bulk of the luminosity arises from thermal processes in the ejecta, which occurs at large distances from the magnetar, where the magnetic fields are not strong enough to produce significant circular polarization, however, such observations are valuable, because they may also allows us to probe for other sources of circular polarization, for example relativistic jets.
\end{enumerate}

\section*{Acknowledgements}

We thank Daniele Malesani for useful discussion. 
This work is on observations made with ESO Telescopes at the Paranal Observatory under the programme ID 098.D-0532(A) and PESSTO (Public ESO Spectroscopic Survey for Transient Objects), ESO programme ID 197.D-1075. SJS acknowledges STFC funding through grant ST/P000312/1. MB acknowledges support from the Swedish Research Council (Vetenskapsr\aa det) and the Swedish National Space Board. TWC acknowledgments the funding provided by the Alexander von Humboldt Foundation.

%%%%%%%%%%%%%%%%%%%%%%%%%%%%%%%%%%%%%%%%%%%%%%%%%%
%%%%%%%%%%%%%%%%%%%% REFERENCES %%%%%%%%%%%%%%%%%%
\bibliographystyle{mnras}
\bibliography{specpol.bib} % if your bibtex file is called example.bib

%%%%%%%%%%%%%%%%%%%%%%%%%%%%%%%%%%%%%%%%%%%%%%%%%%
%%%%%%%%%%%%%%%%% APPENDICES %%%%%%%%%%%%%%%%%%%%%

%\appendix
%\section{Some extra material}

%%%%%%%%%%%%%%%%%%%%%%%%%%%%%%%%%%%%%%%%%%%%%%%%%%

% Don't change these lines
\bsp	% typesetting comment
\label{lastpage}
\end{document}